\documentclass{article}
\usepackage[utf8]{inputenc}
\usepackage{authblk}

\usepackage[english]{babel}
\usepackage{latexsym}
\usepackage{indentfirst,cite}
\usepackage{amsmath}
\usepackage{amssymb}
  \usepackage{pdfpages}
 \usepackage{bm}
 \usepackage{comment}
\usepackage{graphicx} 
\usepackage{amsfonts,soul,mathrsfs}

\newcommand{\be}{\begin{equation}}
 \newcommand{\ee}{\end{equation}}
\newcommand{\bear}{\be\begin{array}}
\newcommand{\bea}{\begin{eqnarray}}
\newcommand{\eea}{\end{eqnarray}}

\newcommand{\bp}{{\bf p}}
\newcommand{\br}{{\bf r}}

\newcommand{\bk}{{\bf k}}

\newcommand{\la}{\langle}
\newcommand{\ra}{\rangle}

\newcommand{\dst}{\displaystyle}
\newcommand{\fr}[2]{\frac{{\dst #1}}{{\dst #2}}}

\begin{document}

\title{Generation of vortex particles via generalized measurements}

\author[$1$]{D.\,V.~Karlovets\thanks{Corresponding author}}
\author[$1$]{S.\,S.~Baturin}
\author[$2$]{G.~Geloni}
\author[$1$]{G.\,K.~Sizykh}
\author[$1,3,4$]{V.\,G.~Serbo}
\affil[$1$]{{\small School of Physics and Engineering, ITMO University, 
197101 St.\,Petersburg, Russia}}
\affil[$2$]{{\small European XFEL, Holzkoppel 4, 22869 Schenefeld, Germany}}
\affil[$3$]{{\small Novosibirsk State University, 630090 Novosibirsk, Russia}}
\affil[$4$]{{\small Sobolev Institute of Mathematics, 630090 Novosibirsk, Russia}}

\date{\today}


\maketitle

\begin{abstract}
The hard X-ray twisted photons and relativistic massive particles with orbital angular momentum -- vortex electrons, muons, protons, etc. -- have many potential applications in high-energy and nuclear physics. However, such states can be obtained so far mainly via diffraction techniques, not applicable for relativistic energies. Here we show that the vortex states of different particles, including hadrons, ions, and nuclei, can be generated in a large class of processes with two final particles simply by altering a postselection protocol. Thanks to entanglement and to the uncertainty relations, an evolved state of a final particle becomes twisted if the momentum azimuthal angle of the other particle is measured with a large uncertainty. We give several examples, including Cherenkov and undulator radiation, particle collisions with intense laser beams, $e\mu \to e\mu, ep \to ep$. This technique can be adapted for ultrarelativistic lepton and hadron beams of linear colliders, and it can also facilitate the development of sources of X-ray and $\gamma$-range twisted photons at storage rings and free-electron lasers.
\end{abstract}

\section{Introduction}

Twisted light with orbital angular momentum (OAM) projection $\la \hat{L}_z\ra \ne 0$ \cite{Allen} has found numerous applications in quantum optics and information, optomechanics, biology, astrophysics, and so forth \cite{Torres, And, SerboUFN, Tamb, Budker, TairaX, Ivanov-atoms}. Along with the diffraction techniques, such photons can be generated by charged particles in undulators \cite{Sasaki1, Sasaki2, Afan, Bahrdt, Kaneyasu, BKL, BKL2}, via non-linear Thomson or Compton scattering \cite{Serbo2011-1, Serbo2011-2, Taira, Katoh, Katoh2, Epp, BKL2}, during Cherenkov and transition radiation \cite{BKLCh}, via channeling in crystals \cite{Chan1, Chan2}, etc. However, despite the potential use in particle and nuclear physics \cite{Budker}, the highest energy of the twisted photons achieved so far does not exceed a few keV \cite{TairaX}.

It has been realized that potentially any quantum wave -- be it an electron \cite{Bliokh}, a hadron \cite{neu}, an ion \cite{Floettmann, NJP}, or a spin wave \cite{Magnon} -- can be created in a twisted quantum state. The vortex electrons, generated at the $300$ keV microscopes \cite{Uchida, Verbeeck, McMorran2}, have attracted much attention outside the microscopy community because of their potential applications in hadronic and spin studies, atomic and high-energy physics, and even in accelerator physics \cite{Bliokh, Ll, Ivanov11, I-S, PRA12, Ivanov_PRA_2012, Ivanov12, Surzh, Serbo15, Kaminer2016, Ivanov2016, Ivanov16, JHEP, Sherwin1,  Sherwin2, PRA18, Ivanov20201, Ivanov20202, Madan, IvanovMu, Floettmann, NJP}. The possible experiments with vortex muons, hadrons, ions, etc. are being discussed \cite{Ivanov_PRA_2012, Ivanov12, JHEP, PRA18, Ivanov20201, Ivanov20202, PRC, Madan, IvanovMu, Floettmann, NJP, Ivanov-atoms} (see the recent review \cite{IvanovRev}), whereas the non-relativistic twisted atoms and molecules have been generated only recently \cite{Atoms}. However, the available diffraction techniques \cite{Uchida, Verbeeck, McMorran2, Beche, neu} are not applicable for relativistic energies, which severely limits the development of the matter waves physics. To probe the vortex physics at higher energies, there is an urgent need in alternative approaches to generate the twisted states of a wide range of quantum systems.

Here, we put forward a method to generate the vortex states of particles of in principle arbitrary mass, spin, and energy, including $\gamma$-rays, relativistic muons, protons, ions and nuclei, 
during the photon emission, scattering, and annihilation processes with two final particles.
The key observation is that it is largely \textit{not the process itself} that defines vorticity of a final particle, but a post-selection protocol due to entanglement between the final particles. Whereas in the classical theory the radiation is twisted if the emitting electron path is helical \cite{Katoh2}, the more general quantum theory developed here predicts that the photons \textit{cannot be twisted at all} within the customary plane-wave postselection \cite{BLP, Peskin} and that the vorticity depends on the way we post-select the electron. That is why the twisted photons may \textit{not} be that abundant in Nature as it seems to follow from the classical theory.

Next, we demonstrate a deep analogy between a so-called \textit{generalized measurement} \cite{BarGen, FrankeGen} of the momentum ${\bm p} = \{p_x, p_x, p_z\}$, in which not all the components are measured with a vanishing uncertainty, and a standard projective (von Neumann) measurement in cylindrical basis \cite{Ivanov_PRA_2012}, in which the azimuthal angle $\phi = \arctan(p_y/p_x)$ is not measured at all. In particular, if the momentum azimuthal angle of a final particle $X$ is measured with a large uncertainty in a process with two final particles $X$ and $Y$, the evolved (pre-selected) state of the other particle $Y$ naturally becomes twisted, thanks to entanglement and to the angle-OAM uncertainty relation \cite{Carruthers, Padgett}. 

Thus, in contrast to the previously employed methods, neither modifications of the incoming beams are required to generate the vortex states nor there are any limitations to the transverse coherence length of the beams; one only needs to postselect one of the final particles differently. This technique clearly demonstrates an advantage of the generalized measurements in the processes of high-energy physics and it can readily be employed for the generation of relativistic vortex beams and of the hard X-ray or $\gamma$-range twisted photons at the electron and hadron accelerators, free-electron lasers and synchrotron radiation facilities, powerful lasers, and at the future linear colliders. A system of units with $\hbar = c = 1$ is used.





\section{Measurement scenarios} 

Let us consider photon emission by an electron, $e \to e' + \gamma$ (e.g., synchrotron radiation, Cherenkov radiation, etc.). An initial state $|\text{in}\ra$ of the electron and an \textit{evolved} (pre-selected) state of the final photon and of the final electron are connected via an evolution operator $\hat{S}$ \cite{BLP, Peskin},
\be
|e', \gamma\ra = \hat{S}\,|\text{in}\ra = \sum\limits_{\lambda_{\gamma}\lambda'} \int\fr{d^3p'}{(2\pi)^3}\fr{d^3k}{(2\pi)^3}\,S_{fi}\,|\bk,\lambda_{\gamma};\bp',\lambda'\ra,
\ee
where $S_{fi} = \la\bk,\lambda_{\gamma};\bp',\lambda'|\hat{S}|\text{in}\ra$ is a transition matrix element with two final plane-wave states with the momenta $\bp', \bk$ and the helicities $\lambda'=\pm 1/2,\lambda_{\gamma} = \pm 1$. Without post-selection, the wave function of the evolved state is \textit{not} factorized into a product of the photon wave function and that of the electron. Indeed, in the momentum representation we have
\bea
\la \bp', \bk |e', \gamma\ra = \sum\limits_{\lambda_{\gamma},\lambda'} u'\, {\bm e}\, S_{fi},
\eea
where ${\bm e} \equiv {\bm e}_{\bk\lambda_{\gamma}}$ is a photon polarization vector in the Coulomb gauge, ${\bm e} \cdot \bk = 0, {\bm e}\cdot {\bm e}^*=1$, and $u' \equiv u_{p'\lambda'}$ is an electron bispinor, normalized as $\bar u' u' = 2 m_e$.

In order to derive the evolved wave function of the photon \textit{alone}, we need to post-select the electron. If it is projected to a state $|e'_{\text{det}}\ra$, the photon wave function in the momentum representation becomes
\bea
& {\bm A}^{(f)}(\bk) = \sum\limits_{\lambda_{\gamma} = \pm 1} {\bm e}\,S_{fi},
\label{wfev}
\eea
where $S_{fi} = \la\bk,\lambda_{\gamma};e'_{\text{det}}|\hat{S}|\text{in}\ra$. 

Let the initial electron be described as a plane-wave state propagating along the $z$ axis (say, during Cherenkov emission) with the momentum $\bp = \{0,0,p\}$. If the electron is post-selected, as usually done, to a plane-wave state with the momentum $\bp'$ and the helicity $\lambda'$, the matrix element $S_{fi}^{(\text{pw})}$ and the photon evolved wave function are both proportional to the $4$-momentum conservation delta-function \cite{BLP, Peskin}
\bea
& S_{fi}^{(\text{pw})} \propto \delta (\bp'_{\perp} + \bk_{\perp}) = \fr{1}{p'_{\perp}}\,\delta (p'_{\perp} - k_{\perp}) \left(\delta(\phi' - (\phi_k -\pi))\Big|_{\phi_k \in [\pi,2\pi]} + \delta(\phi' - (\phi_k +\pi))\Big|_{\phi_k \in [0,\pi]}\right),\cr 
& {\bm A}^{(f)}(\bk) \propto \sum\limits_{\lambda_{\gamma} = \pm 1} {\bm e}\, \delta (\bp'_{\perp} + \bk_{\perp}),
\label{evA}
\eea
so the azimuthal angles of both the final momenta are correlated. For instance, if the electron is detected at a certain azimuthal angle $\phi'$, the photon angle $\phi_k$ is also set to a definite value 
\be
\phi_k = \phi' \pm \pi. 
\ee

Thereby, the photon state is automatically projected to a plane wave \textit{without} an intrinsic OAM and with the infinitely wide OAM spectrum. Indeed, a plane wave $\exp\{i\bk\cdot\br\} = \exp\{ik_{\perp}\rho \cos(\phi_k - \phi_r) + ik_z z\}$ with a finite transverse momentum $k_{\perp}$ is an eigenfunction of the OAM operator $\hat{L}_z = -i\partial/\partial \phi_r$ with the vanishing mean value, $\la \hat{L}_z\ra = 0$, whereas the OAM dispersion $\la \hat{L}_z^2\ra$ can easily be shown to be \textit{infinite} when the wave is normalized in a large cylinder with the volume $V =\pi R^2 L \to \infty$, i.e. the OAM spectrum is flat. Thus, the customary postselection to the plane waves is not suitable for direct comparison with the classical theory from \cite{Sasaki1, Sasaki2, Afan, Taira, Katoh, Katoh2}, 
and this seems to contradict the Bohr correspondence principle.

The plane-wave postselection represents a standard von Neumann (projective) measurement when all components of the momentum $\bp' = \{p'_x,p'_y,p'_z\}$ are measured with vanishing errors. In \textit{a generalized measurement}, some of the components can be measured with a \textit{finite} uncertainty (see, e.g., \cite{BarGen, FrankeGen}). Put simply, a generalized measurement is a postselection to a wave packet,
\bea
|e'_{\text{det}}\ra^{(\text{g})} = \int\fr{d^3p'}{(2\pi)^3}\, \psi(\bp')\, |\bp',\lambda'\ra.
\label{wp}
\eea
with a normalized function $\psi(\bp')$. The set of operators $|{e'_{\text{det}}\ra}^{(\text{g})}\la {e'_{\text{det}}}|^{(\text{g})}$ is complete, although \textit{not necessarily orthogonal},
and it forms a positive operator-valued measure \cite{BarGen}. The function $\psi(\bp')$ can have a Gaussian envelope, $\psi(\bp') \propto \prod\limits_{i=x,y,z}\exp\left\{-(p'_i - \la p_i'\ra)^2/\sigma_i^2\right\}$. A projective measurement implies that $\sigma_x, \sigma_y, \sigma_z \to 0$. Alternatively, one can use the cylindrical coordinates $p'_{\perp}, \phi' = \arctan(p'_y/p'_x), p'_z$ with the uncertainties $\sigma_{\perp}, \sigma_{\phi}, \sigma_z$. For the generalized measurement, at least one of these uncertainties can be \textit{not} vanishing. Let us distinguish the following three measurement scenarios: 

\textit{(i)} We postselect to the plane-wave states \cite{BLP, Peskin} and repeat the measurements many times with an ensemble of electrons, each time fixing the detector at a different azimuthal angle $\phi'$. The emission rate or the scattering cross section are proportional to
\bea
\int\limits_0^{2\pi}\fr{d\phi'}{2\pi}\,|S_{fi}^{\rm(pw)}|^2,
\label{S1}
\eea
which represents \textit{an incoherent} averaging over the azimuthal angle.

\textit{(ii)} Another example of a projective measurement is postselection to a Bessel state \cite{Ivanov_PRA_2012} with the definite $p'_{\perp}, p'_z$, the $z$-projection of \textit{the total angular momentum} (TAM) $m'$, and the helicity $\lambda'$, but undefined $\phi'$ (in accord with the uncertainty relations \cite{Carruthers, Padgett})
\bea
& |e'_{\text{det}}\ra = |p'_{\perp}, p'_z, m', \lambda'\ra = \cr
& = \int\limits_0^{2\pi}\fr{d\phi'}{2\pi}\,i^{-(m'-\lambda')} e^{i(m' - \lambda')\phi'}\,|\bp', \lambda'\ra.
\label{cylexp}
\eea
The corresponding amplitude
\bea
\int\limits_0^{2\pi}\fr{d\phi'}{2\pi}\,i^{m'-\lambda'} e^{-i (m'-\lambda') \phi'}\,S_{fi}^{\rm (pw)}
\label{S2}
\eea
represents \textit{a coherent} averaging over the azimuthal angle, while the detector is able to measure the TAM with a vanishing error $\sigma_{m} \to 0$. As the azimuthal angle and the z-projection of the angular momentum represent \textit{the conjugate variables} \cite{Carruthers, Padgett}, in this way we also obtain complete information about the electron, but in the cylindrical basis.

\textit{(iii)} Consider now an electron emitting a photon when we measure the final electron momentum angle $\phi'$ with an uncertainty close to its maximum value, $\sigma_{\phi} \to 2\pi$. The information about the final electron state is \textit{incomplete}, although the energy is well-defined, $\varepsilon' = \sqrt{(p'_{\perp})^2 + (p'_z)^2 + m_e^2}$. The corresponding amplitude is also obtained via coherent averaging of the plane-wave amplitude, 
\bea
\int\limits_0^{2\pi}\fr{d\phi'}{2\pi}\,S_{fi}^{\rm(pw)}.
\label{S3}
\eea
This expression formally coincides with (\ref{S2}) at $m'-\lambda'=0$, but its physical meaning is different. In the scheme \textit{(ii)}, we do measure the TAM projection and the helicity, and we can easily obtain $\la \hat{L}_z\ra = m'-\lambda'=0$, but during the generalized measurement we do not measure the TAM at all, which implies projection to the state 
\be
|e'_{\text{det}}\ra^{\text{(g)}} = 
\int\limits_0^{2\pi}\fr{d\phi'}{2\pi}\,|\bp', \lambda'\ra,
\label{cyl0}
\ee
where, in contrast to Eq.(\ref{cylexp}) and (\ref{S2}), each plane wave enters with the same phase.
More generally, in this scheme we post-select to a packet (\ref{wp}) with the \textit{finite} uncertainties $\sigma_{\perp}, \sigma_{\phi}, \sigma_z$, 
whereas in Eq.(\ref{cyl0}) we have taken the limiting case $\sigma_{\phi} \to 2\pi$ for simplicity. 

\begin{figure}[h]
	\center
\includegraphics[width=0.9\textwidth]{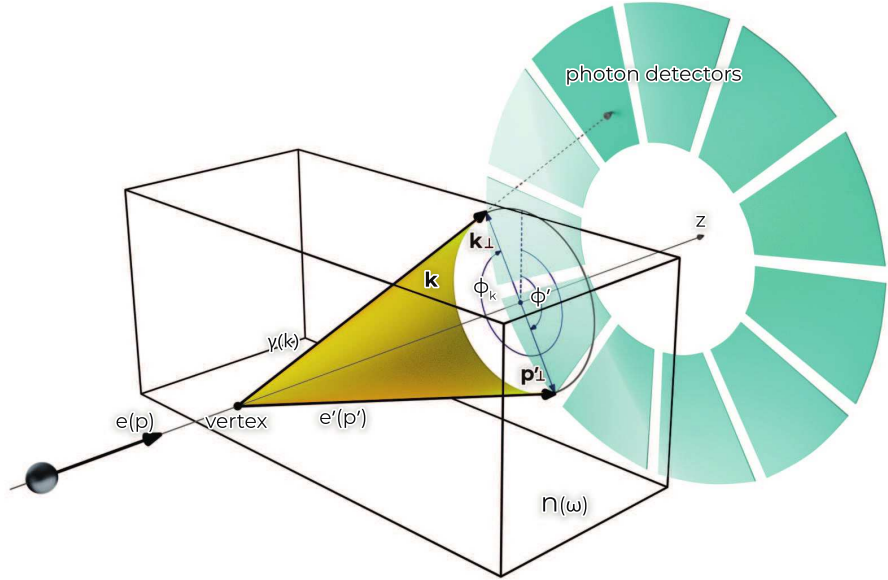}
\caption{An azimuthal ``which-way'' experiment with Cherenkov radiation: during a projective measurement of the electron momentum $\bp'$ we learn the photon angle $\phi_k$, hence, the photon is a plane wave with vanishing OAM. However if the electron angle $\phi'$ is not precisely measured -- say, when the scattering angle is very small, $\theta' \ll 1$ -- the angle $\phi_k$ stays undefined, and the photon evolved state becomes twisted.}
\label{Fig1}
\end{figure}

\section{Cherenkov radiation} 

As an example, consider the radiation, $e(p) \to e'(p') + \gamma(k)$, in a transparent medium with a refractive index $n(\omega)$ by a plane-wave electron with the mass $m_e$ and the following transition amplitude
\bea
& S_{fi}^{\rm(pw)} = -ie N (2\pi)^4 \delta^{(4)}(p-p'-k)\, \bar{u}'\gamma^{\mu}u\, e^*_{\mu},
\eea
where $N$ is a normalization constant, $\gamma^{\mu}$ are the Dirac matrices \cite{BLP} and
\bea
& p = \{\varepsilon, 0, 0,|\bm p|\},\ p' = \{\varepsilon', p'_{\perp}\cos\phi', p'_{\perp}\sin\phi', p'_z\},\cr 
& p'_{\perp} = |\bm p'| \sin \theta',\ k = \{\omega, k_{\perp} \cos\phi_k, k_{\perp} \sin\phi_k, k_z\},\cr 
& |\bk| = \sqrt{k_{\perp}^2 + k_z^2} = \omega\, n(\omega).
\eea

The azimuthal angles of the final momenta are correlated during the customary projective measurements, and thus, we know which detector in Fig.\ref{Fig1} registers the photon. If the electron is detected in the above generalized measurement scheme (\textit{iii}), we do not know the photon azimuthal angle, and its evolved state is obtained by putting Eq.(\ref{S3}) to (\ref{wfev}) as
\bea
& {\bm A}^{(f)}(\bk) = \int\limits_0^{2\pi}\fr{d\phi'}{2\pi} \sum\limits_{\lambda_{\gamma} = \pm 1} {\bm e}\, S_{fi}^{\rm(pw)} =  \cr 
& = - ieN\, {\bm n}\times\left[{\bm n}\times \int\limits_0^{2\pi}\fr{d\phi'}{2\pi} (2\pi)^4 \delta^{(4)}(p-p'-k)\, \bar{u}'{\bm \gamma}u\right],
\label{Phcyl}
\eea
where we employ the Coulomb gauge with $e^{\mu} = \{0,{\bm e}\}$, $\sum\limits_{\lambda_{\gamma}=\pm 1} e_i e^*_j = \delta_{ij} - n_i n_j,\ {\bm n} = \bk/|\bk|$. 
We choose the overall phases of the electron bispinors $u$ so that $\hat{j}_z u = \lambda u,\, \hat{j}_z= \hat{s}_z + \hat{L}_z$ (as in Ref.\cite{Peskin}).
The electron transition current $\bar{u}'{\bm \gamma}u$ is found as
\bea
& \bar{u}'{\bm \gamma}u = \left(\sqrt{\varepsilon' + m_e}\sqrt{\varepsilon - m_e} + 2\lambda\,2\lambda'\sqrt{\varepsilon + m_e}\sqrt{\varepsilon' - m_e}\right)\cr 
& \times \left(d^{(1/2)}_{\lambda\lambda'}(\theta')\, {\bm \chi}_{0}  - \sqrt{2}\,d^{(1/2)}_{-\lambda\lambda'}(\theta')\, {\bm \chi}_{2\lambda}\, e^{-i 2\lambda \phi'} \right)\, e^{i (\lambda - \lambda')\phi'},
\label{curr}
\eea
where $\lambda'$ is the final electron helicity, ${\bm \chi_0} = (0,0,1)^T, {\bm \chi}_{\pm 1} = \mp \frac{1}{\sqrt{2}}(1,\pm i, 0)^T$ are eigenvectors of the spin operator $\hat{s}_z$ with the eigenvalues $0, \pm 1$, and the small Wigner functions are 
\be
d_{\lambda\lambda'}^{(1/2)}(\theta') = \delta_{\lambda\lambda'} \cos(\theta'/2) - 2 \lambda\,\delta_{\lambda,-\lambda'} \sin(\theta'/2). 
\ee

The final expression for the photon wave function is
\bea
& {\bm A}^{(f)}(\bk) = (-1)^{\lambda-\lambda'} i e N (2\pi)^3 \delta(\varepsilon - \varepsilon' -\omega) 
\delta (|\bm p| - p_z' - k_z)\, \fr{1}{p'_{\perp}}\, \delta (p'_{\perp} - k_{\perp})\cr
& \times \Big(\sqrt{\varepsilon' + m_e}\sqrt{\varepsilon - m_e} + 2\lambda\,2\lambda'\sqrt{\varepsilon + m_e}\sqrt{\varepsilon' - m_e}\Big)
\left[\bm F - \bm n (\bm n \bm F)\right],\cr 
& \bm F =d_{\lambda\lambda'}^{(1/2)}(\theta')\, {\bm \chi}_{0}\, e^{i(\lambda - \lambda')\phi_k} + \sqrt{2}\,d_{-\lambda\lambda'}^{(1/2)}(\theta')\,{\bm \chi}_{2\lambda} e^{-i(\lambda + \lambda')\phi_k}.
\label{PolA}
\eea
The terms in the vector $\bm F$ are eigenvectors of $\hat{s}_z$ operator with the eigenvalues $0$ and $2\lambda$ and of the OAM operator $\hat L_z=-i \partial/\partial \phi_k$ with the eigenvalues $\lambda-\lambda'$ and $-\lambda-\lambda'$, respectively. Therefore, the photon evolved state is twisted
\bea
& \hat j_z^{(\gamma)}\bm A^{(f)} =(\lambda-\lambda') \bm A^{(f)},\ \hat{j}^{(\gamma)}_z = \hat{s}_z + \hat{L}_z,\cr
& j_z^{(\gamma)} = \lambda-\lambda' = -1, 0, \text{or}\, +1,
\label{TAMcl1}
\eea
where even the state with $j_z^{(\gamma)} = 0$ is not a plane wave, but a twisted one.
The transverse momentum $k_{\perp} = |\bm p'| \sin \theta'$ of this \textit{Bessel beam} with the spin-orbit interaction (SOI) is defined by the electron scattering angle $\theta'$. As this angle is very small, $\theta' \ll 1$, it is technically challenging to precisely measure the azimuthal angle $\phi'$, and in this case the Cherenkov photons become naturally twisted.


When the initial electron is \textit{twisted} itself with the TAM $m = \pm 1/2, \pm 3/2, ...$ \cite{Bliokh}, its bispinor $u \equiv u_{p \lambda}$ with $\bm p = \{p_\perp\cos\phi, p_\perp\sin\phi,p_z\}$ transforms as (see Eq.~\eqref{cylexp}) 
\be
u_{p\lambda} \to u_{p_\perp p_z m \lambda}=
i^{-(m-\lambda)}\int\limits_0^{2\pi} \fr{d\phi}{2\pi} 
\,e^{i(m-\lambda)\phi}\,u_{p \lambda}.
 \label{u-twisted}
\ee
As a result, the photon TAM becomes 
\be
\lambda- \lambda' \to j_z^{(\gamma)} = m-\lambda',
\ee
and $|j_z^{(\gamma)}|$ can be larger than $1$ for the vortex electrons with $m \gg 1$. Thus, generating Cherenkov radiation via highly twisted electrons and combining this with the generalized measurement of the electron azimuthal angle results in the highly twisted photons. 

\section{Non-linear Compton scattering and undulator radiation} 

In a circularly polarized laser wave with the potential \cite{BLP, R}
\bea
& A^{\mu} = a_1^{\mu} \cos(kx) + a_2^{\mu} \sin(kx),\, kx = \omega t - \bk \cdot \br, \cr
& A^2 = a_1^2 = a_2^2 = -a^2 < 0,\, (a_1 a_2) = 0,
\eea
an electron is described with a Volkov state \cite{BLP, R, Piazza, Fedotov}
\bea
&\psi_{p\lambda}(\br,t) = N_e \left(1 + \fr{e}{2(pk)} (\gamma k)(\gamma A)\right)\cr 
&\times u\, \exp\left\{-ipx - \fr{ie}{(pk)} \int\limits^{kx}d\varphi \left((pA) - \fr{e}{2}A^2\right)\right\}.
\eea
The final photon wave function is a plane wave ${\mathcal A}^{\mu} = N_{\gamma}\, e'^{\mu}\, e^{-i\omega't + i\bk'\cdot\br}$. The matrix element is
\bea
S_{fi}^{(\text{pw})} = -ie\int d^4 x\, \bar{\psi}_{p'\lambda'}\gamma_{\mu}\psi_{p\lambda} \left({\mathcal A}^{\mu}\right)^*
\eea
where the final electron is also in the Volkov state.

The results of this exactly solvable problem can be applied for an approximate description of similar problems where an electron also moves along a helical path, the simplest examples being emission in a helical undulator or in a longitudinal magnetic field. The characteristics of radiation in a helical undulator and in the above laser wave are quantitatively very similar for ultrarelativistic electrons with $\varepsilon/m \gg 1$ and small recoil $\omega'/\varepsilon \ll 1$ \cite{UndCompt}. Next, the electron is usually postselected to the Volkov state \cite{BLP, R, Piazza, Fedotov}, which implies a projective measurement of its quasi-momentum $q'$ with the vanishing uncertainties. This may \textit{not} necessarily happen in the quasi-classical regime when the recoil is small, the laser pulse (or the undulator) is long so the electron emits several photons and stays inside the field after emission, but the angle $\phi'$ may not be precisely measured after each emission event. So we do not get complete information about the final electron state and in this regime the photon can naturally become twisted. 


Following the standard procedure \cite{BLP, R, Piazza, Fedotov}, we expand the matrix element into a series over the harmonic number $s = 1,2,3,...$, collect the terms with the same indices 
of the Bessel functions $J_s$ and $J_{s\pm 1}$, and get
\bea
& S_{fi}^{(\text{pw})} = \sum\limits_{s=1}^{\infty} S_{fi}^{(s)} = -ie N \sum\limits_{s=1}^{\infty} (2\pi)^4 \delta^{(4)} (q + sk - q'- k')\cr 
& \times \sum\limits_{\sigma = 0, \pm 1} (e'_{\mu})^*\,\bar{u}'\,\Gamma_{\sigma}^{\mu}\,u\, J_{s+\sigma}(\Sigma)\, e^{i (s + \sigma)\xi},
\label{SVolkov}
\eea
which is a standard expression, just written differently. The ``dressed'' vertex $\Gamma_{\sigma}^{\mu}$ is
\bea
& \Gamma_{\sigma}^{\mu} = \{\Gamma_{0}^{\mu}, \Gamma_{+1}^{\mu}, \Gamma_{-1}^{\mu}\} = \cr
& = \Big\{\gamma^{\mu} - k^{\mu} \fr{e^2 A^2}{2(pk)(p'k)} (\gamma k), \fr{1}{2}(\gamma a_-) \left (\fr{e}{2} (\gamma k) \gamma^{\mu} (\fr{1}{(p'k)} - \fr{1}{(pk)}) + k^{\mu} \fr{e}{(pk)}\right) - a_-^{\mu}\fr{e}{2(pk)} (\gamma k),\cr & \fr{1}{2}(\gamma a_+) \left (\fr{e}{2} (\gamma k) \gamma^{\mu} (\fr{1}{(p'k)} - \fr{1}{(pk)}) + k^{\mu} \fr{e}{(pk)}\right) - a_+^{\mu}\fr{e}{2(pk)} (\gamma k)\Big\},\cr
& a_{\pm}^{\mu} = a_1^{\mu} \pm i a_2^{\mu}, (ka_{\pm}) = 0,
\label{VVertex}
\eea
and other notations are
\bea
& q^{\mu} = p^{\mu} + e^2\,k^{\mu}\fr{a^2}{2(pk)},\, (q')^{\mu} = (p')^{\mu} + e^2\,k^{\mu}\fr{a^2}{2(p'k)}\cr
& \Sigma^2 = e^2\left(\fr{(pa_1)}{(pk)} - \fr{(p'a_1)}{(p'k)}\right)^2 + e^2\left(\fr{(pa_2)}{(pk)} - \fr{(p'a_2)}{(p'k)}\right)^2,\cr
& \xi = \arctan\fr{(pa_2)/(pk) - (p'a_2)/(p'k)}{(pa_1)/(pk) - (p'a_1)/(p'k)}.
\eea
We only study the head-on collision with $\xi = \phi' = \phi_{k'} \pm \pi$ where $\phi_{k'}$ is the angle of the final photon momentum $\bk' = \omega' {\bm n}'$.

After the summation over helicities, the evolved state of the photon {\it at the $s$th harmonic} within the generalized-measurement scheme becomes (more detailed mathematical derivations will be presented elsewhere)
\bea
& \displaystyle {\bm A}^{(f,s)}_{(\text{g})}(\bk') = \int\limits_0^{2\pi}\fr{d\phi'}{2\pi}\, {\bm A}^{(f,s)}(\bk') = -ie (-1)^{s+\lambda-\lambda'} N (2\pi)^3 \delta (q^0 + s\omega - (q')^0- \omega') \delta (q_z + sk_z - q_z'- k_z')\cr 
& \displaystyle \times \fr{1}{k_{\perp}'}\delta (p_{\perp}' - k_{\perp}')\, {\bm n}' \times \left[{\bm n}' \times \sum\limits_{\sigma=0,\pm 1} J_{s+\sigma}(\rho'_e k'_{\perp})\,
e^{i (s + \sigma - \lambda')\phi_{k'}} \left(d^{(1/2)}_{\lambda\lambda'}\,{\bm G}^{(\uparrow \uparrow)}\, e^{i\lambda\phi_{k'}} +
d^{(1/2)}_{-\lambda\lambda'}\,{\bm G}^{(\uparrow \downarrow)}\, e^{-i\lambda\phi_{k'}}\right)\right],
\label{Astr}
\eea
where we have denoted 
\bea
&& {\bm G}^{(\uparrow \uparrow)}_{0\lambda'\lambda} = {\bm \chi}_{0} \left( f^{(2)}_{\lambda'\lambda} - \fr{\eta^2 m_e^2 \omega^2}{2(pk)(p'k)} (f^{(1)}_{\lambda'\lambda} + f^{(2)}_{\lambda'\lambda})\right),\cr
&& {\bm G}^{(\uparrow \downarrow)}_{0\lambda'\lambda} = - f^{(2)}_{\lambda'\lambda} \sqrt{2}\, {\bm \chi}_{2\lambda},\cr
&& {\bm G}^{(\uparrow \uparrow)}_{\pm 1\lambda'\lambda} = \mp\sqrt{2}\, {\bm \chi}_{\mp 1} \fr{\eta m_e \omega}{2} \left(f^{(1)}_{\lambda'\lambda} + f^{(2)}_{\lambda'\lambda}\right) \left(\delta_{\lambda,\mp 1/2} \left(\fr{1}{(p'k)} - \fr{1}{(pk)}\right) + \fr{1}{(pk)}\right),\cr
&& {\bm G}^{(\uparrow \downarrow)}_{\pm 1\lambda'\lambda} = \mp {\bm \chi}_{0} \fr{\eta m_e \omega}{2}\delta_{\lambda,\pm 1/2} \left(\left(\fr{1}{(p'k)} - \fr{1}{(pk)}\right) (f^{(1)}_{\lambda'\lambda} - f^{(2)}_{\lambda'\lambda}) - \fr{2f^{(2)}_{\lambda'\lambda}}{(pk)}\right),\cr
&& f^{(1)}_{\lambda'\lambda} = \sqrt{\varepsilon + m_e} \sqrt{\varepsilon' + m_e} + 2\lambda 2\lambda' \sqrt{\varepsilon - m_e} \sqrt{\varepsilon' - m_e},\cr
&& f^{(2)}_{\lambda'\lambda} = \sqrt{\varepsilon - m_e} \sqrt{\varepsilon' + m_e} + 2\lambda 2\lambda' \sqrt{\varepsilon + m_e} \sqrt{\varepsilon' - m_e},
\label{GG}
\eea
and $\eta = e\sqrt{-A^2}/m_e$ is a classical field strength parameter \cite{BLP, Piazza, Fedotov}.
Eq.(\ref{Astr}) describes \textit{a Bessel beam} with the SOI and the following TAM projection (cf. Eq.(\ref{TAMcl1})):
\bea
\hat{j}_z^{(\gamma)} {\bm A}^{(f,s)}_{(\text{g})} = (s + \lambda - \lambda') {\bm A}^{(f,s)}_{(\text{g})}.
\eea

When the incoming electron is twisted with the TAM $m$ (being in the Bessel-Volkov state \cite{PRA12}), we obtain 
\bea
s + \lambda - \lambda' \to j_z^{(\gamma)} = s + m - \lambda', 
\eea
whereas 
for the unpolarized electrons the photon TAM is simply $s$, which is in agreement with the classical theory \cite{Sasaki1, Sasaki2, Afan, Taira, Katoh, Katoh2}. 
Thus, the correspondence principle is clearly demonstrated within the generalized measurement scheme but not with the projective measurements in the plane-wave basis.

The above quantum theory clarifies the conditions required to generate hard X-ray or $\gamma$-range twisted photons by relativistic charged particles colliding with laser beams or moving in helical undulators. Namely, the evolved state of photons becomes naturally twisted only when we do \textit{not} measure the azimuthal angle of the charged particle momentum. An alternative approach \textit{(ii)} from Sec.2 above -- used, for instance, in \cite{BKL, BKL2, BKLCh} -- would be to apply the standard von Neumann measurements by using a special detector that projects the evolved photon state to the twisted one. Clearly, the development of such a detector can represent a separate experimental challenge, especially for highly energetic particles, and we will not discuss this problem here.

\section{Heavier leptons, hadrons, and nuclei} 

We take the process
 \be
e^{-}(p_1)+\mu^{-}(p_2) \to e^{-}(p_3) +\mu^{-}(p_4) 
 \ee
as an example, where the electron and the muon have the momenta $p_1 = \{\varepsilon_1, 0, 0,|\bm p_1|\}, p_2 = \{\varepsilon_2 , 0, 0, -|\bm p_2|\}$
and the helicities $\lambda_1, \lambda_2$, respectively. When the electron is detected in the above scheme \textit{(iii)}, the evolved wave function of the muon becomes
\be
\psi^{(f)}_\mu = \sum\limits_{\lambda_4 = \pm 1/2} 
\int\limits_0^{2\pi} \fr{d\phi_3}{2\pi}\,
u_4\, S_{fi}^{\rm(pw)},
\label{psimu}
\ee
where $u_4 \equiv u_{p_4\lambda_4}$ and the matrix element reads
\bea
& S_{fi}^{\rm(pw)} = i(2\pi)^4 N\, \delta^{(4)}(p_1 + p_2 - p_3 - p_4)\cr 
& \times \fr{4\pi e^2}{q^2}\, \left(\bar{u}_3 \gamma^{\alpha}u_1\right)\, \left(\bar{u}_4 \gamma_{\alpha}u_2\right),\, q = p_1 - p_3.
\eea
Here, the propagator taken in the Feynman gauge and the muon transition current $\bar{u}_4 \gamma_{\alpha} u_2$ can be presented analogously to Eq.(\ref{curr}).

\begin{figure}[h]
\center
\includegraphics[width=0.9\textwidth]{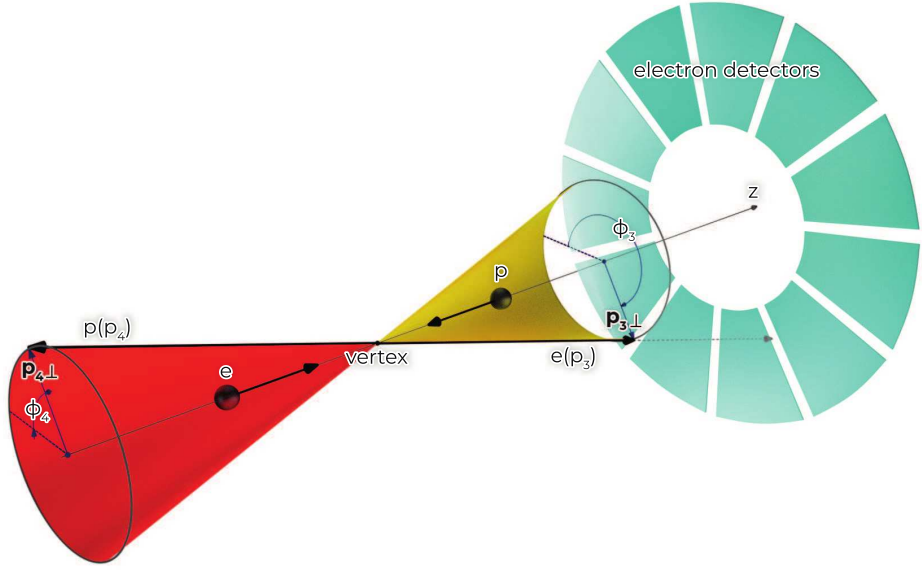}
\caption{An azimuthal ``which-way'' experiment with elastic $ep \to ep$ scattering. The proton becomes twisted if the electron azimuthal angle $\phi_3$ is measured with a large uncertainty $\sigma_{\phi} \to 2\pi$ or is not measured at all. Instead of the proton, any other hadron, ion, or nucleus can be used.}
\label{Fig2}
\end{figure}
 
After the integration in Eq.(\ref{psimu}), we arrive at 
 \bea
 \psi^{(f)}_\mu \propto \sum\limits_{\lambda_4 = \pm 1/2} \tilde{u}_4\,\, \mathcal J\, e^{i(\lambda_1-\lambda_2-\lambda_3)\phi_4},
 \eea
where $\tilde{u}_4 \equiv u_4\,e^{- i \lambda_4 \phi_4}$ has a vanishing TAM, $\hat{j}_z \tilde{u}_4 = 0$ (as in Ref.\cite{BLP}), and the factor $\mathcal J$ does not depend on $\phi_4$. 
As a result, the muon evolved state is proved to be twisted,
 \be
{\hat j}_{4,z}\psi^{(f)}_\mu= (\lambda_1-\lambda_2-\lambda_3)\psi^{(f)}_\mu.
\ee

Finally, we examine scattering off a proton (see Fig.\ref{Fig2}),
\be 
e(p_1) + p(p_2) \to e(p_3) + p(p_4),
\ee
in the same head-on geometry. Generalization to other hadrons, nuclei, or to inelastic processes is straightforward.
The matrix element is
\bea
& S_{fi}^{\rm(pw)} = i(2\pi)^4 N\, \delta^{(4)}(p_1 + p_2 - p_3 - p_4)\cr 
& \times \fr{4\pi e^2}{q^2}\, \left(\bar{u}_3 \gamma^{\mu}u_1\right)\, \left(\bar{u}_4 \Gamma_{\mu}u_2\right),
\eea
where $\Gamma_{\mu} = F_1 \, \gamma_{\mu} + F_2 \, \sigma_{\mu\nu}q^{\nu}$ is a hadronic vertex, represented via the form-factors $F_1 = F_1 (q^2, P^2), F_2 = F_2 (q^2, P^2)$ \cite{BLP, Peskin}, where $q^2 = (p_1 - p_3)^2,\, P^2 = (p_4 + p_2)^2/4$ do not depend on the angle $\phi_4$ of the final proton.

Similar to above, one can prove that 
\be
\bar{u}_3(\phi_3 = \phi_4 \pm \pi)\gamma^{\mu} u\,
\left(\bar{u}_4 \Gamma_{\mu}u_2\right)
\propto e^{i (\lambda_1 - \lambda_2 - \lambda_3 - \lambda_4)\phi_4}
\ee
and within the same generalized-measurement protocol the evolved state of the proton becomes twisted,
\be 
\hat j_{4,z} \psi^{(f)}_p= (\lambda_1-\lambda_2-\lambda_3)\,
  \psi^{(f)}_p
\ee 
or 
\bea
\lambda_1-\lambda_2-\lambda_3 \to m-\lambda_2-\lambda_3 
\eea
when the initial electron is also twisted with the TAM $m$.
For the unpolarized particles, the electron TAM can be transferred in this way to a hadron, pion, ion, or a nucleus. 

\section{Conclusion} 

We have shown that the vortex quantum states of highly energetic photons, of relativistic leptons and hadrons, including ions and nuclei, can be generated in a large family of customary scattering, radiation, or annihilation processes simply by employing the postselection protocol with a generalized measurement of the momentum azimuthal angle. In particular, this scheme imposes no limitations on the transverse coherence length of the incoming beams, and it adequately describes the emission of twisted photons in the classical regime. Moreover, the proposed method deals with the particle states as they have evolved from the process themselves and it does not invoke to any kind of special detector. Clearly, the ways how one can probe vorticity of the final particles depend on the particle mass, charge, and energy, and we will not discuss these means here. One can envisage the implementation of this technique, for instance, at the SASE3 undulator beamline of the European XFEL, at such powerful laser facilities as the Extreme Light Infrastructure, at synchrotrons with the helical undulators, and at the existing and future lepton and hadron colliders.

\

We are grateful to A.~Di Piazza, A.~Tishchenko, A.~Surzhykov, A.~Pupasov-Maksimov, and A.~Volotka for fruitful discussions and criticism. The studies on the measurement schemes to generate twisted photons are supported by the Russian Science Foundation (Project No. 21-42-04412) and by the Deutsche Forschungsgemeinschaft (Project No. SU 658/5–1). The studies on the Compton scattering and undulator radiation are supported by the Ministry of Science and Higher Education of the Russian Federation (agreement no. 075-15-2021-1349). The studies on heavy particles are supported by the Government of the Russian Federation through the ITMO Fellowship and Professorship Program. The work on the evolved states of particles (by D.~Karlovets and G.~Sizykh) was supported by the Foundation for the Advancement of Theoretical Physics and Mathematics “BASIS”.


\end{document}